\documentstyle[aps,prl,psfig,floats]{revtex}

\begin{document}
\draft
\title{A deterministic origin for instabilities in a magneto-optical trap}
\author{Andrea di Stefano, Marie Fauquembergue, Philippe Verkerk and Daniel Hennequin}
\address{Laboratoire de Physique des Lasers, Atomes et Mol\'{e}cules, UMR CNRS,
Centre d'Etudes et de Recherches Lasers et Applications, Universit\'{e} des
Sciences et Technologies de Lille, F-59655 Villeneuve d'Ascq cedex - France}
\date{\today }
\twocolumn[\hsize\textwidth\columnwidth\hsize\csname @twocolumnfalse\endcsname
\maketitle

\begin{abstract}
The paper reports on the observation of a new type of instabilities in the
atomic cloud of a magneto-optical trap (MOT). They appear as large amplitude
self-oscillations, either periodic or erratic, depending on the parameters.
An original model taking into account the shadow effect and an homogeneous
distribution of the atoms in the cloud is developped. It allows us to show
that these instabilities originate on the one hand from an anomalous Hopf
bifurcation, and on the other hand from the extreme sensitivity of the
system to noise.
\end{abstract}


\vskip2pc] %valider pour prl

\renewcommand\floatpagefraction{.9} %valider pour prl
\renewcommand\topfraction{.9} %valider pour prl
\renewcommand\bottomfraction{.9} %valider pour prl
\renewcommand\textfraction{.1} %valider pour prl

The magneto-optical cooling of atoms is at the origin of a renew of the
atomic physics. It is used in various fields, as the realization of
Bose-Einstein condensates\cite{BEC} or the study of diffusion in lattices 
\cite{lattices}, and could lead to several applications, as atomic clocks 
\cite{clock} or quantum computing\cite{IQ}. Although the technology and
realization of magneto-optical traps (MOT) is well mastered, some
experimental adjustments remain empirical. It is in particular well known
that for dense atomic clouds close from resonance, instabilities appear in
the spatio-temporal distribution of the atoms. This problem is usually fixed
by empirically misaligning the trapping beams.

A recent study has shown that instabilities can have a stochastic origin,
through a phenomenon of coherent resonance amplifying the technical noise 
\cite{david}. It also showed that the shadow effect is at the origin of
these instabilities : because of the absorption of light inside the cloud,
the intensities of the backward and forward beams are locally different,
leading to an internal attracting force. In the configuration where each
backward beam is obtained by retro-reflection of the forward beam, backward
and forward intensities are globally different, and an external force
appears, displacing the cloud along the bisectors of the trap beams.

In the present work, we show that the atomic cloud in the MOT can also
exhibit deterministic periodic and chaotic-like instabilities. The basic
model developed in \cite{david} is not able to reproduce them, and we
propose a modified model where the cloud is no more considered as pointlike.
This new model predicts the same type of stationary solutions as in \cite
{david}, leading to the same type of stochastic instabilities. Moreover, it
allows us to retrieve the deterministic instabilities: in particular, it
exhibits large amplitude periodic oscillations induced by an anomalous Hopf
bifurcation, and a large amplitude stochastic behavior. It also establish
definitively the role of the shadow effect in the dynamics of the atomic
cloud in the MOT.

The initial hypotheses are the same as in \cite{david}, i.e.\ a 1D model
taking into account the shadow effect. The system is modelized on the one
hand through the motion equations of the center of mass $z$ of the cloud,
and on the other hand through a rate equation of the number of atoms $n$ in
the cloud. We have:

\begin{mathletters}
\label{ModInit}
\begin{eqnarray}
\frac{d^{2}z}{dt^{2}} &=&\frac{1}{M}F_{T} \\
\frac{dn}{dt} &=&B\left( n_{e}-n\right)
\end{eqnarray}
where $M$ is the mass of the cloud, $F_{T}$ the total external force, $n_{e}$
the atom number at equilibrium and $B$ the population relaxation rate. As in 
\cite{david}, $n_{e}$ is assumed to depend on $z$, to take into account the
depopulation of the cloud when it moves away from the trap center. We define
a distance $z_{0}$ beyond which the trap is empty ($n_{e}=0$). For $z<z_{0}$%
, we assume a quadratic behavior $n_{e}=n_{0}\left( 1-\left( z/z_{0}\right)
^{2}\right) $ where $n_{0}$ is the cloud population at the trap center \cite
{david}.

The innovation of this model concerns the way the shadow effect is treated.
We consider that, starting from an input forward intensity $I_{1}$, the
intensity after a first crossing of the cloud (i.e. the input backward
intensity) is $I_{2}<I_{1}$, and the remaining intensity after a second
crossing of the cloud (i.e. the output backward intensity) is $I_{3}<I_{2}$.
The rate of photons absorbed in the forward (resp. backward) beam is $%
A\left( I_{1}-I_{2}\right) $ (resp. $A\left( I_{2}-I_{3}\right) $), with $%
A=S/h\nu $, where $S$ is the section of the cloud and $h\nu $ the energy of
a photon. The forces associated with each beam is the product of the number
of absorbed photons by the elementary momentum $\hbar k$. We obtain finally: 
\end{mathletters}
\begin{equation}
F_{T}=\frac{S}{c}\left( I_{1}-2I_{2}+I_{3}\right)  \label{FT}
\end{equation}

To get a relation between $I_{1}$, $I_{2}$, and $I_{3}$, we need to solve
the equations of propagation of the two beams through the atomic cloud. For
the sake of simplicity, we assume that the atomic transition is a $%
J=0\rightarrow J=1$ transition. As the MOT is operated with high intensity
beams and small detunings, a Doppler model is suited. Inside the cloud, the
intensity $I_{+}$ ($I_{-}$) of the $\sigma _{+}$ ($\sigma _{-}$) polarized
forward (backward) beam evolves due to photon scattering. The scattering
rate for $\sigma _{\pm }$ photons is proportional to the corresponding
excited state populations $\Pi _{\pm }$. The evolution equations of the
intensity simply writes 
\begin{equation}
\frac{dI_{\pm }}{dz}=\mp \Gamma h\nu \rho \Pi _{\pm }  \label{evolint}
\end{equation}
where $\Gamma $ is the natural width of the atomic transition and $\rho $ is
the atomic density in the cloud. The populations $\Pi _{\pm }$ are given by
the steady state of the master equation. The underlying hypothesis is that
the evolution of the external degrees of freedom is much slower than that of
the internal ones. The populations $\Pi _{\pm }$ depend both on $I_{+}$ and $%
I_{-}$, so that (\ref{evolint}) is a set of coupled nonlinear equations.
They are integrated numerically from the side of the cloud where $%
I_{+}=I_{-}=I_{2}$, to the other side, where $I_{-}=I_{3}$ and $I_{+}=I_{1}$%
, assuming that the density $\rho $ is constant, as a result of multiple
scattering \cite{multi}.

The control parameters of this system are the same as in \cite{david}. The
most suitable one is the detuning $\Delta _{0}$, expressed in units of $%
\Gamma $, between the trap laser beams and the atomic transition. The global
organization of stationary solutions in the parameter space is also the same
as in \cite{david}. In particular, stationary solutions as a function of $%
\Delta _{0}$ evolve from a flat dependence to bistability when another
parameter (e.g. $n_{0}$) is changed. The fold appearing for the intermediate
values is at the origin of the stochastic instabilities reported in \cite
{david}.

\begin{figure}[tph]
\centerline{\psfig{figure=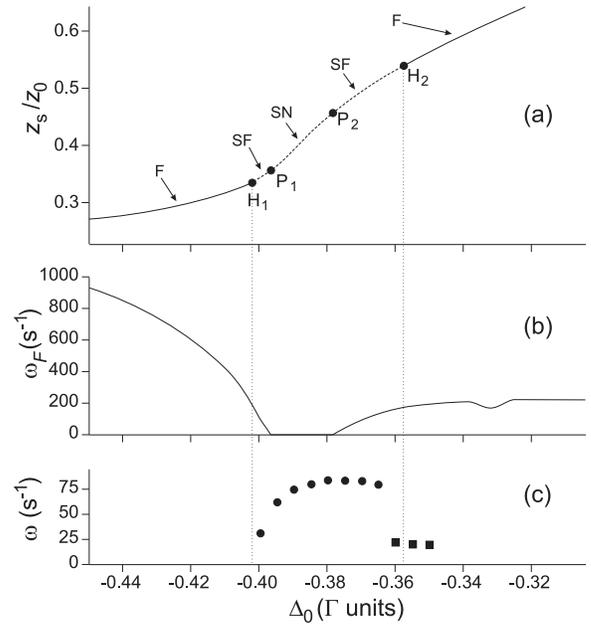,width=3in}}
\caption{Theoretical evolution of the behavior of the cloud as a function of
the detuning. In (a), the stationary solution $z_s$ of $z$ is stable (full
line) or unstable (dashed line). In points H$_1$ and H$_2$, a Hopf
bifurcation occurs, while in points P$_1$ and P$_2$, $\protect\omega_F$
vanishes. F (focus), SF (saddle focus) and SN (saddle node) refer to the
nature of the fixed point representing the stationary solution in the phase
space. In (b), evolution of $\protect\omega_F$ versus $\Delta_0$. In (c),
plot of the instability frequencies $\protect\omega_A$ (circles) and $%
\protect\omega_B$ (squares). Parameters for the calculations are $I_1=33$
mW/cm$^2$, $\protect\rho=2\times 10^{10}$ cm$^{-3}$, $n_0=6\times 10^8$, $%
z_0=3$ cm, $B=5$ s$^{-1}$ and a Zeeman shift of $3\Gamma$ cm$^{-1}$.}
\label{fig:fig1}
\end{figure}

We focus here on the unstable zone appearing between the area studied in 
\cite{david} and bistability. In this situation, the stationary solutions
are unstable on the fold (fig. \ref{fig:fig1}). For detuning smaller than
the fold, at the left of point H$_{1}$ on fig. \ref{fig:fig1}a, the fixed
point is a stable focus (F): the stationary solutions are stable and
associated with an eigenfrequency $\omega _{F}$ decreasing with the detuning
(fig. \ref{fig:fig1}b). At the edge of the fold, the system exhibits a Hopf
bifurcation (point H$_{1}$): the fixed point becomes a saddle-focus (SF),
and the stationary solutions become unstable. As the detuning is further
increased, the eigenvalues become real in point P$_{1}$ (fig. \ref{fig:fig1}%
a,b), so that $\omega _{F}$ disappears and the fixed point becomes a saddle
node (SN). Finally, when the detuning is still increased, the inverse
sequence appears for the fixed point (SN $\rightarrow $ SF $\rightarrow $
Hopf bifurcation $\rightarrow $ F).

Let us now examine the dynamical behavior of the atomic cloud in the
different regions. In the F zone, the stationary solution is stable. As the
detuning is increased, the stationary solution becomes unstable in H$_{1}$,
and a stable periodic orbit $C_{H}$ appears in the vicinity of the fixed
point, as it is usual with a Hopf bifurcation. The atomic cloud exhibits
weak amplitude oscillations around the unstable fixed point, with a
frequency close to $\omega _{F}$. The orbit amplitude increases slowly with
the detuning, but remains very weak (less than 10 nm in $z$ for the
parameters of fig. \ref{fig:fig1}).

\begin{figure}[tph]
\centerline{\psfig{figure=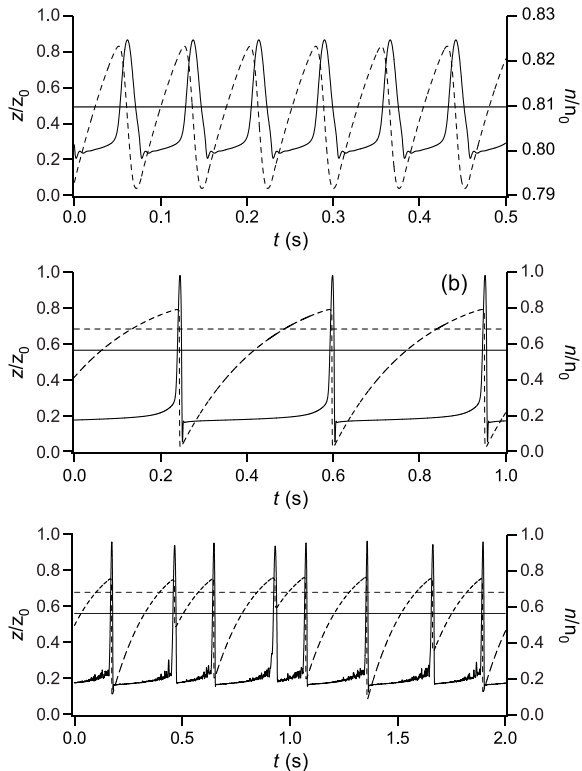,width=3in}}
\caption{Examples of the behavior of the cloud. The full (dashed) line curve
is a plot of $z$ ($n$) versus time. The horizontal full (dashed) line marks
the stationary value $z_s$ ($n_s$). In (a), $n_s/n_0=0.757$ is outside the
figure. (a) shows a C$_A$ instability for $\Delta_0=-0.37$ ; (b) shows a C$%
_B $ instability for $\Delta_0=-0.35$; (c) corresponds to the same
parameters as (b), but a noise level of $7\%$ has been added on $I_1$. Other
parameters are the same as in fig. \ref{fig:fig1}.}
\label{fig:fig2}
\end{figure}

In fact, if $\Delta _{0}$ is further increased, $C_{H}$ disappears for a
value $\Delta _{A1}$ of $\Delta _{0}$ very close from H$_{1}$. It is
replaced by another limit cycle $C_{A}$, which differs from $C_{H}$ on three
main points: its amplitude is much larger than that of $C_{H}$, its
frequency $\omega _{1}$ is not $\omega _{F}$, and its shape is not
sinusoidal at all. Fig. \ref{fig:fig2}a shows an example of a $C_{A}$ cycle:
in each period, a slow increasing of $z$ is followed by a fast growth where $%
z$ becomes much larger than $z_{s}$, and a fast decreasing after which the
cycle starts again. Meanwhile, $n$ increases slowly and regularly, then
decreases. Note that in the example of fig. \ref{fig:fig2}a, the stationary
value $n_{s}$ of the population is outside the limit cycle, and so the
interpretation of the behavior in terms of the relative values of $z$ and $n$
with respect to $z_{s}$ and $n_{s}$ is not straightforward. In particular,
it cannot be interpreted as a relaxation oscillation process around the
unstable stationary values. In fact, as we show below, this behavior is not
connected with the unstable fixed point properties.

Because $C_{H}$ becomes unstable for a value so close from H$_{1}$, the Hopf
bifurcation appears as anomalous: while such a bifurcation leads usually to
progressively growing cycle, it generates here a large amplitude periodic
orbit. However, the $C_{A}$ behavior is not linked to the SF zone: it still
exists beyond P$_{1}$ and P$_{2}$, without any discontinuity. The amplitude
of $C_{A}$ is several millimeters when it appears in $\Delta _{A1}$, and
increases regularly with the detuning, so that for a value $\Delta _{A2}$ of 
$\Delta _{0}$, located between P$_{2}$ and H$_{2}$ for the parameters of
fig. \ref{fig:fig1}, the cloud border reaches $z_{0}$. At this point, the
shape of the limit cycle qualitatively changes. Indeed, the atoms beyond $%
z_{0}$ are lost, and so the population in the cloud can decrease rapidly.
The resulting temporal behavior is still a periodic cycle $C_{B}$ and may be
described as previously, except that the decreasing of $z$ is much faster
and that of the population much larger (fig. \ref{fig:fig2}b). Note that
this behavior is observed beyond H$_{2}$: this means that generalized
bistability occurs between $C_{B}$ and the stable stationary solution. This
confirms that the periodic instabilities are not directly linked to the
fixed point properties.

The $C_{H}$, $C_{A}$ and $C_{B}$ behaviors differ also by their frequency.
As discussed above, the $C_{H}$ frequency is $\omega _{F}$, and so varies
rapidly as a function of the detuning. The frequency $\omega _{A}$ of the $%
C_{A}$ regime is very small when the regime appears in $\Delta _{A1}$ (fig. 
\ref{fig:fig1}c). When the detuning is increased, $\omega _{A}$ reaches
rapidly a value of the order of $2\pi \times 10$ s$^{-1}$, and then remains
almost constant. Finally, the frequency $\omega _{B}$ of the $C_{B}$ regime
is also almost constant when $\Delta _{0}$ is changed, with a frequency
smaller than $\omega _{A}$ (about $\omega _{A}/4$ in fig. \ref{fig:fig1}).

Previous studies have shown the fundamental role of noise in this system 
\cite{david}. The influence of noise on deterministic instabilities is well
known: fixed points and limit cycles are usually robust with respect to
noise, whose main effect is to slightly shift the bifurcation points \cite
{noise}. So we do not expect to observe spectacular changes in the
stationary, C$_{H}$ and C$_{A}$ behaviors if noise is added in the system,
and this is confirmed by the simulations. The C$_{B}$ behavior is different,
as the cloud could be very sensitive to noise in the vicinity of $z_{0}$:
indeed, noise should induce large variations in the decreasing of $n$, and
hence in the period of the dynamics. This is confirmed by the numerical
simulations: fig. \ref{fig:fig2}c shows the behavior of the cloud for the
same conditions as in fig. \ref{fig:fig2}b, except that noise has been added
on the trap intensity. As expected, the dynamics is no more periodic,
exhibiting large fluctuations in the return time of the dynamics.

To check the existence of the deterministic instabilities predicted by the
model, we used the same experimental setup as described in \cite{david},
i.e. a three-arm $\sigma ^{+}-\sigma ^{-}$ magneto-optical trap (MOT) on
Cesium, with mirrors to produce the counter-propagating waves of each arm,
and a crossed couple of 4-quadrant photodiodes to monitor the cloud. As in 
\cite{david}, forward and backward beams are carefully aligned. The only
changes with respect to the experiment described in \cite{david} concern the
parameter values, and in particular the intensity of the trap laser beams,
adjusted to obtain deterministic instabilities.

\begin{figure}[tph]
\centerline{\psfig{figure=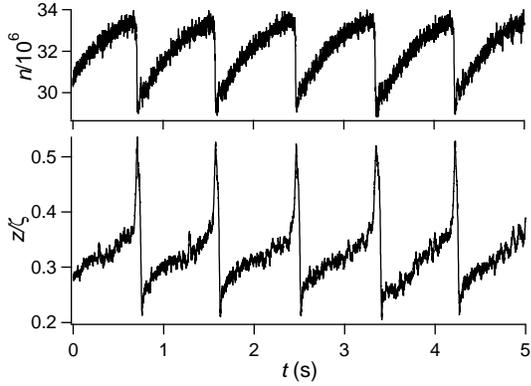,width=3in}}
\caption{Experimental record of a C$_A$ periodic instability. Parameters are 
$I_1=11$ mW/cm$^2$ and $\Delta_0=-1.4$. The beam waist of the trap beams is
5 mm, while the magnetic field gradient is 13 G/cm. $\protect\zeta$ is
proportionnal to the mean size of the cloud.}
\label{fig:fig3}
\end{figure}

\begin{figure}[tph]
\centerline{\psfig{figure=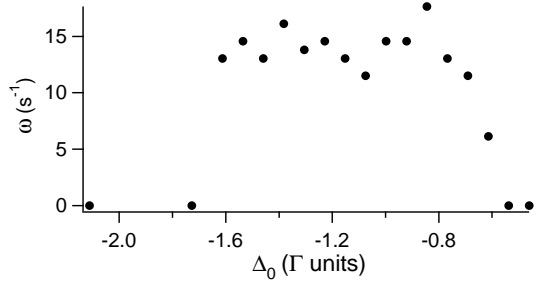,width=3in}}
\caption{Evolution of the instability frequency $\protect\omega$ versus the
detuning. Parameters are the same as in Fig. \ref{fig:fig3} except $I_1=20$
mW/cm$^2$.}
\label{fig:fig4}
\end{figure}

As expected, the $C_{H}$ cycle is not observed. The most commonly observed
periodic instabilities look as $C_{A}$ instabilities (fig. \ref{fig:fig3}).
Fig. \ref{fig:fig4} shows the evolution of the signal frequency $\omega $ as
a function of the detuning. As in the model, $\omega $ is almost constant on
the interval where $C_{A}$ instabilities appear, i.e. for $-1.7<\Delta
_{0}<-0.8$. The value of $\omega $ depends on the experimental parameters,
as the trap beam or the repump intensities. It is typically of the order of
one or a few hertz, in good agreement with the model. On the contrary, the
detuning interval on which instabilities appear is one order of magnitude
larger than that predicted. This is the main discrepancy between the
experimental observations and the model. However, to make a real comparison,
we should take into account the inevitable variation of $n_{0}$ when $\Delta
_{0}$ is changed. Note that in the model, a simultaneous change of $n_{0}$
and $\Delta _{0}$ leads to a relative stretch of the unstable zone.
Unfortunately, as we have no way to establish experimentally the relation
between $n_{0}$ and $\Delta _{0}$, we are not able to check the amplitude of
the correction in the present model.

\begin{figure}[tph]
\centerline{\psfig{figure=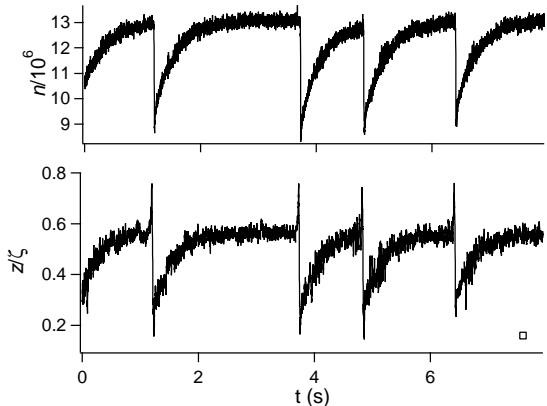,width=3in}}
\caption{Experimental record of a C$_2$-like instability. Parameters are the
same as in Fig. \ref{fig:fig4} with $\Delta_0=-0.6$.}
\label{fig:fig5}
\end{figure}

When the detuning is increased, the behavior shifts towards $C_{B}$
instabilities. Because of noise, the transition is not abrupt as in the
simulations. The limit cycle becomes more and more noisy as $\Delta _{0}$ is
increased, while the frequency decreases. Finally, for $\Delta _{0}>-0.55$,
the instabilities disappear and the behavior is again stationary. As
expected in the above discussion, the $C_{B}$ instabilities appear in their
erratic form (fig. \ref{fig:fig5}). The main characteristics of the
experimental behavior corresponds to the theoretical results: the decreasing
stages of $z$ and $n$ are very fast, while the mean frequency of the signal
is shorter than that of $C_{A}$.

In conclusion, we have demonstrated the existence of a new type of
instabilities in the MOT clouds. These instabilities, contrary to those
already observed, are deterministic, implying that a simple amelioration of
the experimental noise cannot improve the cloud stability. Instabilities
consist in large amplitude periodic or erratic cycles. We have proposed a
new model taking into account the intensity distribution inside the cloud.
This model allows us to interprete the instabilities through an anomalous
Hopf bifurcation generating the large amplitude pulses, and shows that even
in the case of deterministic instabilities, noise can play a main role in
the behavior. These results are obtained for a three-beam configuration, but
we are expecting them to be more general. In particular, in the six beam
configuration, the shadow effect also induces nonlinearities, leading to
similar dynamics on the atomic density. However, the present model is still
too simple to give a quantitative agreement with the experiments. A 3D model
with a non constant atomic density will give a finer description of the
dynamics.

The Laboratoire de Physique des Lasers, Atomes et Mol\'{e}cules is
``Unit\'{e} Mixte de Recherche de l'Universit\'{e} de Lille 1 et du CNRS''
(UMR 8523). The Centre d'Etudes et de Recherches Lasers et Applications
(CERLA) is supported by the Minist\`{e}re charg\'{e} de la Recherche, the
R\'{e}gion Nord-Pas de Calais and the Fonds Europ\'{e}en de
D\'{e}veloppement Economique des R\'{e}gions.

% now the references. delete or change fake bibitem. delete next three
% lines and directly read in your .bbl file if you use bibtex.


%\begin{table}[tbp]
%\caption{}
%\label{}
%\begin{tabular}{|c|c|}
%\end{tabular}
%\end{table}


\begin{references}
\bibitem{BEC}  F. Dalfovo et al, Rev. Mod. Phys. {\bf 71}, 463 (1999)

\bibitem{lattices}  C. Jurczak et al, Phys. Rev. Lett. {\bf 77}, 1727
(1996); L. Guidoni et al, Phys. Rev. A {\bf 60}, R4233 (1999)

\bibitem{clock}  see e.g. Y. Sortais et al, Phys. Rev. Lett. {\bf 85}, 3117
(2000) and references therein.

\bibitem{IQ}  G. K. Brennen et al, Phys. Rev. Lett. {\bf 82}, 1060 (1999)

\bibitem{david}  D. Wilkowski et al, Phys. Rev. Lett. {\bf 85}, 1839 (2000)

\bibitem{multi}  D. W. Sesko et al, J. Opt. Soc. Am. B {\bf 8}, 946 (1991)

\bibitem{noise}  E. Arimondo et al, in {\em Noise in Nonlinear Dynamical
Systems, vol III}, F. Moss \& P. V. E. McClintock eds, Cambridge University
Press (1989)
\end{references}
\end{document}